\documentstyle[12pt]{article}
\newcommand{\be}{\begin{equation}}
\newcommand{\ee}{\end{equation}}
\newcommand{\ba}{\begin{eqnarray}}
\newcommand{\ea}{\end{eqnarray}}

\begin{document}
\begin{center}
{\bf PSEUDO-HERMITICITY OF AN EXACTLY SOLVABLE TWO-DIMENSIONAL MODEL
}\\
\vspace{0.5cm} {\large \bf F. Cannata$^{1,}$\footnote{E-mail:
cannata@bo.infn.it}, M.V. Ioffe$^{2,}$\footnote{E-mail:
m.ioffe@pobox.spbu.ru} ,
D.N. Nishnianidze$^{2,3,}$\footnote{E-mail: qutaisi@hotmail.com}}\\
\vspace{0.2cm}
$^1$ Dipartimento di Fisica and INFN, Via Irnerio 46, 40126 Bologna, Italy.\\
$^2$ Department of Theoretical Physics, Sankt-Petersburg State University,\\
198504 Sankt-Petersburg, Russia\\
$^3$ Akaki Tsereteli State University, 4600 Kutaisi, Georgia
\end{center}
\vspace{0.2cm} \hspace*{0.5in}
%\vspace{1cm}
\hspace*{0.5in}
\begin{minipage}{5.0in}
{\small We study a two-dimensional exactly solvable non-Hermitian $PT-$non-symmetric quantum model with real
spectrum, which is not amenable to separation of variables, by supersymmetrical methods. Here we focus
attention on the property of pseudo-Hermiticity, biorthogonal expansion and pseudo-metric operator.
To our knowledge this is the first time that pseudo-Hermiticity is realized explicitly for a
nontrivial two-dimensional case. It is shown that the Hamiltonian of the model is not diagonalizable.\\
\vspace*{0.1cm} PACS numbers: 03.65.-w, 03.65.Fd, 11.30.Pb }
\end{minipage}
\vspace*{0.2cm}
\section{\bf Introduction.}
\vspace*{0.1cm} \hspace*{3ex} Supersymmetrical techniques has been
successfully applied to two-dimensional Quantum Mechanics (see \cite{david} and
the review paper \cite{ioffe}). For the Hermitian case several {\bf
real} two-dimensional models - Morse potential \cite{new}, P\"oschl-Teller potential
\cite{iv} and some others \cite{innn} - were studied
by means of two different SUSY methods: SUSY-separation
of variables and shape invariance. Some partial solutions of the
spectral problems were obtained by this approach.

During the last decade intensive study of Schr\"odinger equation with
complex potentials, but with real spectrum, was performed by different methods.
The pioneer papers \cite{bender1} initiated investigation of $PT-$symmetric systems
(see also the review papers \cite{bender2}),
and afterwards more general class of pseudo-Hermitian models was
considered \cite{mostafazadeh}.

One key tool for the {\bf complexification}
of two-dimensional models with real spectra is given by the intertwining relations between
partner Hamiltonians with supercharges of second order in
derivatives. A particular class of models - complex {\bf singular}
two-dimensional Morse - has been found \cite{pseudo} to satisfy
SUSY-pseudo-Hermiticity, i.e.
\be
 H^{\dagger}Q^+ = Q^+H, \label{intertw}
\ee
where the complex supercharges intervene in the HSUSY
deformation of the standard algebra of SUSY QM. Only partial
knowledge of the spectrum and wave functions for this model was
obtained.

More recently another complexification of the {\bf real} singular
Morse model was considered \cite{exact}, which from now on we will
call {\bf regularized} complex Morse system. This Hamiltonian
being also involved in second order SUSY intertwining {\bf does not}
satisfy (\ref{intertw}), but we point out that it fulfills
standard pseudo-Hermiticity as defined in \cite{mostafazadeh}:
\be
H^{\dagger} = \eta H \eta^{-1} \label{pseudo}
\ee
with some
invertible positive-definite operator $\eta .$ In particular,
since in this model the complexification arises from a complex
coordinate shift, which also provides a regularization,
pseudo-Hermiticity is rather straightforward \cite{ahmed}. Due to
the fact that this model turns out to be solvable, we can now
discuss explicitly the biorthogonal expansion based on the eigenfunctions
and their complex conjugated.

The structure of the paper is the following. The main results concerning the partially
solvable model with real two-dimensional Morse potential are reproduced in Section 2.
In Section 3 the exactly solvable two-dimensional regularized complex Morse model is
analysed. This model is not $PT-$symmetric, but its energy eigenvalues are real.
After the description of its spectrum and corresponding eigenfunctions, the
biorthogonal basis based on these eigenfunctions and their complex conjugated is
studied. The explicit construction indicates that the Hamiltonian is actually not
diagonalizable (concerning one-dimensional non-diagonalizable Hamiltonians see,
for example, \cite{non} and references therein),
and biorthogonal basis has to be completed in order to provide the
resolution of identity. Finally, according to the well known prescriptions, the
pseudo-metric operator $\eta$ and the corresponding positively definite (pseudo)inner
product \cite{mostafazadeh}, \cite{tanaka} space are displayed explicitly.

\section{\bf The partially solvable real two-dimensional Morse potential.}
\vspace*{0.1cm} \hspace*{3ex} The pseudo-Hermitian model which we
want to study in the next Section originates from the following
Hermitian Hamiltonians:
\ba
\tilde H(\vec x) &=& - \Delta +
\frac{\alpha^2a(2a-1)}{\sinh^{2}\biggl(\frac{\alpha
x_-}{2}\biggr)}  + A \biggl[e^{-2\alpha x_1}-2 e^{-\alpha x_1} +
e^{-2\alpha x_2}-2 e^{-\alpha x_2}\biggr] + 4a^2\alpha^2, \label{tildemorse}\\
H(\vec x) &=& - \Delta +
\frac{\alpha^2a(2a+1)}{\sinh^{2}\biggl(\frac{\alpha
x_-}{2}\biggr)} + A \biggl[e^{-2\alpha x_1}-2 e^{-\alpha x_1} +
e^{-2\alpha x_2}-2 e^{-\alpha x_2}\biggr] + 4a^2\alpha^2,
\label{morse}
\ea
where the parameters $a, A, \alpha > 0$ are
arbitrary real numbers, and $x_{\pm}\equiv x_1 \pm x_2$.

For the particular choice of parameter $a=-1/2,$ SUSY intertwining
relations
\be
Q^+H=\tilde H Q^+;\quad HQ^-=Q^-\tilde H \label{intertww}
\ee
with the supercharges:
\ba
Q^{\pm}&=&4\partial_+\partial_- \mp 2\alpha\partial_- \mp 2\alpha
\coth\frac{\alpha x_-}{2}\partial_+ + \alpha^2\coth\frac{\alpha
x_-}{2}-\nonumber\\&-& A\biggl[e^{-2\alpha x_1}-2 e^{-\alpha x_1}
- e^{-2\alpha x_2}+2 e^{-\alpha x_2}\biggr];\quad \partial_{\pm}=
\frac{\partial}{\partial x_{\pm}}
\label{superch}
\ea
link the Hermitian Hamiltonian $\tilde H(\vec x)$
from (\ref{tildemorse}) to a partner $ H(\vec x)$ of (\ref{morse})
which does not contain the first term in the r.h.s of
(\ref{morse}) and therefore allows for separation of variables
\cite{exact}. The latter Hamiltonian $ H(\vec x)$ is
straightforwardly solvable with energies expressed in terms of two
integer positive numbers:
\be
E_{n,m}=\epsilon_n+\epsilon_m+\alpha^2;\quad \epsilon_k \equiv
-A[1-\frac{\alpha}{\sqrt{A}}(k+1/2)]^2;\quad k,n,m = 0,1,\ldots
 \label{energy}
\ee
All levels of $H$ with $n\neq m$ are two-fold degenerated and the corresponding wave
functions can be chosen as symmetric and antisymmetric combinations:
\be
\Psi^S_{n,m}=\Psi_{n,m}+\Psi_{m,n};\qquad \Psi^A_{n,m}=\Psi_{n,m}-\Psi_{m,n},
\label{AS}
\ee
where the functions $\Psi_{n,m}$ were defined as:
\be
\Psi_{n,m} =\eta_n(x_1)\eta_m(x_2) .\label{psinm}
\ee
In turn, $\eta_k$ are
the standard solutions of the one-dimensional Morse problem and
can be written in terms of confluent hypergeometric functions:
\ba
&&\Biggl(-\partial^2 + A \biggl(\exp(-2\alpha x)-2 \exp(-\alpha
x)\biggr)\Biggr)\eta_n(x)=\epsilon_n\eta_n(x);
\label{etta}\\
&&\eta_n = \exp(-\frac{\xi}{2}) (\xi)^{s_n}
\Phi(-n, 2s_n +1; \xi);\quad \xi\equiv \frac{2\sqrt{A}}{\alpha}\exp(-\alpha x);
\label{hyper}\\
&& s_n=\frac{\sqrt{A}}{\alpha}-n-1/2 > 0; \quad \epsilon_n=-\alpha^2s_n^2;\quad n=0,1,\ldots \label{eppsilon}
\ea

The wave functions of $\tilde H(\vec x)$ with the same energies
(\ref{energy}) are obtained from (\ref{AS}) acting by
supercharge $Q^+$ from (\ref{superch}):
\be
\tilde\Psi^{A(S)}_{n,m}=Q^+\Psi^{S(A)}_{n,m}. \label{Q+}
\ee
The operator $Q^+$ has singular coefficient functions, and it
is antisymmetric under $x_1\Leftrightarrow x_2.$

The two-fold degeneracy of levels (\ref{energy}) of $H$
under $n \leftrightarrow m$ is not reproduced, in general, in the spectrum of $\tilde
H.$
While the
singularities of $Q^+$ at $x_-=0$ can be compensated by $\Psi^A_{n,m}$ for
$\tilde\Psi^S_{n,m}(\vec x)$, the wave functions $\tilde\Psi^A_{n,m}$ may be
nonnormalizable. Up to now the hypergeometric functions in
expressions for the wave functions (\ref{psinm}) {\bf did not allow} to
perform a comprehensive analysis of the normalizability of {\bf
all} wave functions, i.e. to prove the exact solvability of the model (see details and
some examples in \cite{exact}).

\section{\bf The regularized complex Morse model.}
\vspace*{0.1cm}
\hspace*{3ex} In order to avoid the singularities at $x_-\to 0,$
which hinder the solvability of the model, it is useful
\cite{exact} to perform a suitable complex coordinate shift
\be
\vec x\to \vec x + i\vec\delta;\quad \vec\delta=(\delta, 0)
\label{shift}
\ee
with $\delta$ small enough (such that
$\alpha\delta\in (0, \pi/2))$ in order to remove the singularities from the
real $(x_1,x_2)$ plane, preserving the normalizability of the functions
$\eta_n(x_1)$ from (\ref{hyper}) at $x_1\to -\infty$. After this complex shift the Hamiltonian
has obviously a real spectrum, but the analysis of
normalizability of wave functions is now essentially simplified by the absence of
singularities in the supercharges. Complexification of both operators $Q^{\pm}$
is achieved by the same shift (\ref{shift}) in definitions (\ref{superch}).
Therefore, their mutual Hermitian conjugacy is replaced now by
\be
Q^-=((Q^+)^{\dagger})^{\star}.
\label{transp}
\ee
The spectrum of the complexified Hamiltonian $H(\vec x
+i\vec\delta),$ which is still amenable to separation of
variables, coincides with (\ref{energy}), and {\bf all}
eigenfunctions $\Psi_{n,m}$ are obtained from (\ref{psinm}) by the
same imaginary shift of $\vec x.$

Similarly to the Hermitian case, the intertwining relations (\ref{intertww})
lead to the eigenfunctions
$\tilde\Psi_{n,m}(\vec x +i\vec\delta)$ of the non-separable
non-Hermitian Hamiltonian $\tilde H(\vec x + i\vec\delta):$
\be
\tilde\Psi^{A(S)}_{n,m}(\vec x +i\vec\delta)=Q^+(\vec x+
i\vec\delta)\Psi^{S(A)}_{n,m}(\vec x +i\vec\delta), \label{tildepsi}
\ee
but now, due to the absence of singularity of $Q^+$ at $x_-\to 0$,
these wave functions are normalizable.
The corresponding eigenvalues (see (\ref{energy})) are two-fold degenerate: one can choose symmetric or
antisymmetric combinations of $\Psi_{n,m}$. The only exclusions are the levels
$E_{n,n\pm 1},$ which are not degenerate, because antisymmetric functions
$\Psi^A_{n,n\pm 1},$ being \cite{exact} the linear combinations of zero modes of $Q^+,$
are annihilated by $Q^+.$

It is known \cite{david} that both Hamiltonians $H$ and $\tilde H$ obey the dynamical
symmetry properties. The
fourth order operators $R=Q^-Q^+$ and $\tilde R=Q^+Q^-$ commute with $H$ and $\tilde
H$, respectively, while they do not mix the degenerate wave functions. For the case of complex
potentials these operators are not Hermitian because of the relation (\ref{transp}).

In next Section we will need the eigenvalues $r_{n,m}$ of $R$ for eigenfunctions
$\Psi^{S(A)}_{n,m}(\vec x +i\vec\delta).$ They can be
calculated in terms of "one-dimensional energies" $\epsilon_n, \epsilon_m$ of (\ref{energy}).
Indeed, separation of variables in operator $H$ with $a=-1/2$ gives:
\ba
H(\vec x + i\vec\delta)&=&h_1(x_1 + i\delta) + h_2(x_2)+\alpha^2;\label{Hhh}\\
h_{1}&=&-\partial_1^2-f_1=-\partial_1^2+A\biggl(e^{-2\alpha (x_1+i\delta)}-2e^{-\alpha
(x_1+i\delta)}\biggr);\label{h1}\\
h_{2}&=&-\partial_2^2+f_2=-\partial_2^2+A\biggl(e^{-2\alpha x_2}-2e^{-\alpha
x_2}\biggr);\label{h2}
\ea
The explicit form (\ref{superch}) of the supercharges $Q^{\pm}$ leads to the following
expression:
\be
R=Q^-Q^+=\biggl(h_2-h_1+\frac{1}{4}C_+C_--C_+\partial_--C_-\partial_+ \biggr)
\biggl(h_2-h_1+\frac{1}{4}C_+C_-+C_+\partial_-+C_-\partial_+ \biggr),
\ee
which for $a=-\frac{1}{2}$ can be transformed by straightforward calculations to:
\be
R=(h_1-h_2)^2+2\alpha^2(h_1+h_2) +\alpha^4.
\label{Rhh}
\ee
It means that the eigenvalues $r_{n,m}$ are expressed as:
\be
r_{n,m}=(\epsilon_n-\epsilon_m)^2+2\alpha^2(\epsilon_n+\epsilon_m)+\alpha^4=
\alpha^4\biggl((m-n)^2-1\biggr)\biggl((s_m+s_n)^2-1\biggr),
\label{repsilon}
\ee
where the positive parameters $s_n$ were defined in (\ref{eppsilon}). One can notice
that for some integer $n,m$ eigenvalues $r_{n,m}$ are not positive
(operator $R$ is not Hermitian):
$r_{n,n}=\alpha^4(1-4s_n^2)<0$
for all $n$ (excluding $n=[\frac{\sqrt{A}}{\alpha}-\frac{1}{2}]$),
and $r_{n,n\pm 1}=0$ for all values of $n.$

In general, besides eigenfunctions of the form (\ref{tildepsi}) some {\it additional}
normalizable eigenstates of $\tilde H$ could exist,
if they would be annihilated by $Q^-,$ or if they would be transformed by $Q^-$ into nonnormalizable
functions. The second option
is excluded due to nonsingular form of supercharges. The analysis of zero modes
of $Q^-$ is performed analogously to investigation in \cite{new} (Subsections 4.3 - 4.5) but up
to some appropriate
changes in that paper: $Q^+\to Q^-;\,\,h \to \tilde h$ etc\footnote{In particular, it means that
one has to use in these calculations $a=1/2.$}. The required set of $\tilde\Psi_n$ - linear combinations of
$N$ zero modes $\Omega_l;\,\,l=0,1,\ldots ,N$ of $Q^-$ - is constructed
by means of SUSY-separation of variables \cite{new} and the similarity transformation with function
$\xi_1\xi_2(\xi_2-\xi_1)^{-1}:$
\be
\tilde\Psi_{n} = \sum^N_{k=0}b_{nl}\Omega_l,
\label{4}
\ee
where $b_{nl}$ are matrix elements of $\hat B,$ which satisfy the matrix equation:
\be
\hat E \hat B = \hat B \hat C .
\label{B}
\ee
In this equation $\hat E$ is diagonal matrix with elements
\be
E_n=c_{nn}=-2\alpha^2s_n(1+s_n);\quad n=0,1,2,\ldots ,N,
\label{En}
\ee
and $\hat C$ is the triangular matrix \cite{new} with elements $c_{nk},$ such that
\be
\tilde H\Omega_n=\sum^N_{k=0}c_{nk}\Omega_k.
\label{3}
\ee
The direct algorithm for calculation of $b_{nl}$ in terms of known $c_{nk}$ was also given in \cite{new}.

One can notice that the eigenvalues $E_n$ from (\ref{En}) for the values $n=1,2,\ldots ,N$ coincide with
the eigenvalues $E_{n-1,n}$ of $\tilde\Psi_{n-1,n}=Q^+\Psi^S_{n-1,n}$ from (\ref{tildepsi}),
which were found by using intertwining relations.
It is necessary now to compare the corresponding eigenfunctions $\tilde\Psi_n$ and $\tilde\Psi_{n-1,n}.$

Since the eigenvalues $r_{n-1,n}$ of $R=Q^-Q^+$ vanish for all $n=1,2,\ldots ,N,$ the eigenfunctions
$\tilde\Psi_{n-1,n}$ of $\tilde H$ are simultaneously the zero modes of $Q^-$, and therefore
must be linear combinations of $\Omega_k$ with some unknown coefficients $a_{nk}$:
\be
Q^+\Psi^S_{n-1,n}=\sum^N_{k=0}a_{nk}\Omega_{k};\quad n=1,2,\ldots ,N.
\label{7}
\ee
Acting with the $\tilde H$ onto both sides of this relation and subsequently equating coefficients in
front of $\Omega_l$ gives:
\be
E_{n-1,n}a_{nl}=\sum^N_{k=0}a_{nk}c_{kl};\quad n,l=1,2,\ldots ,N.
\label{8}
\ee
In matrix form this equation coincides with (\ref{B}) up to replacing $b_{nk}$ by $a_{nk}$, therefore
these matrix elements also coincide up to a common constant factor.
This analysis shows that
functions $Q^+\Psi^S_{n-1,n}$ coincide with $\tilde\Psi_n$ for $n=1,2,\ldots N,$ and the eigenvalues
$E_{n-1,n}=E_n$ still are not degenerate.

There is {\it only one} additional eigenstate in the spectrum of $\tilde H$
not obtained from intertwining relations. It corresponds to $n=0,$ i.e. $E_0=-2\alpha^2s_0(1+s_0).$
Its wave function - the lowest zero mode of $Q^-$ -  reads:
\be
\tilde\Psi_0=exp(-\frac{\xi_1+\xi_2}{2})(\xi_1\xi_2)^{s_0+1}(\xi_2-\xi_1)^{-1}.
\label{add}
\ee

Thus, the spectrum of the Hamiltonian $\tilde H(\vec x + i\vec\delta)$ is known:
it includes two-fold degenerate levels $E_{n,m}$ with $m\neq n\pm 1,$ non-degenerate
levels $E_{n-1,n}$ with $n=1,2,\ldots ,N$ and one additional level with energy $E_0.$

\section{\bf Biorthogonal basis and pseudo-Hermiticity.}
\vspace*{0.1cm}
\hspace*{3ex} The wave functions $\Psi^{S(A)}_{n,m}(\vec x+i\vec\delta)$ of $H(\vec
x+i\vec\delta)$ (with separation of variables) and their complex
conjugate functions $(\Psi^{S(A)}_{n,m}(\vec x
+i\vec\delta))^{\star}$ form the so called biorthogonal basis for
the non-Hermitian Hamiltonian $H$. The corresponding
biorthogonality relations
\ba
&&\langle\Psi^{\star}_{n,m}\mid
\Psi_{n',m'}\rangle = \int d^2x \Psi_{n,m}(\vec x+i\vec\delta)
\Psi_{n',m'}(\vec x+i\vec\delta)
=\nonumber\\
&&=\int dx_1 \phi_n(x_1+i\delta) \phi_{n'}(x_1+i\delta) \int dx_2
\phi_n(x_2) \phi_{n'}(x_2)=\delta_{nn'}\delta_{mm'} \label{ortho}
\ea
can be checked straightforwardly and by comparing the integral along the line
$x_1\in (-\infty +i\delta, +\infty + i\delta)$ with the integral along the real axis
(with no singularities of integrand between these lines).

The construction of the bound-states-biorthogonal basis by means of the wave functions
$\tilde\Psi_{n,m}(\vec x +i\vec\delta)$ and $\tilde\Psi_0(\vec x+i\vec\delta)$
of $\tilde H(\vec x+i\vec\delta)$ together with
$\biggl(\tilde\Psi_{n,m}(\vec x +i\vec\delta)\biggr)^{\star}$ and
$(\tilde\Psi_0(\vec x+i\vec\delta))^{\star}$ is much less simple.

Due to the property (\ref{transp}), for the complex model the scalar products
analogous to (\ref{ortho}) can be written as:
\ba
 &&\langle\biggl(\tilde\Psi_{n,m}(\vec x+i\vec\delta)\biggr)^{\star}\mid
\tilde\Psi_{n',m'}(\vec x+i\vec\delta)\rangle =\nonumber\\
 &&=\langle (Q^+)^{\star}\biggl(\Psi_{n,m}(\vec x+i\vec\delta)\biggr)^{\star}
 \mid Q^+\Psi_{n',m'}(\vec x+i\vec\delta)\rangle =\nonumber\\
 &&=\langle\Psi^{\star}_{n,m}(\vec x+i\vec\delta)\mid Q^-Q^+
 \Psi_{n',m'}(\vec x+i\vec\delta)\rangle = r_{n,m}\delta_{n,n'}\delta_{m,m'}.
\label{orthoo}
\ea
In the last equality we used the fact that wave
functions $\Psi_{n,m}$ are the common eigenfunctions both of
the Hamiltonian $H$ and of its symmetry operator $R=Q^-Q^+$ with
the real eigenvalues $r_{n,m}.$

For all pairs $n,m$ with $m\neq n\pm 1$
the functions in (\ref{orthoo}) can be made orthonormal by suitable normalization
factors, real or imaginary depending on the sign of $r_{n,m}.$
In particular, for $m=n$ one can choose  $i|r_{n,n}|^{-1/2}\tilde\Psi_{n,n}(\vec
x+i\vec\delta)$ and $\biggl(i|r_{n,n}|^{-1/2}\tilde\Psi_{n,n}(\vec
x+i\vec\delta)\biggr)^{\star}$ as the elements of biorthogonal basis $(r_{n,n}<0)$.

No analogous simple prescription works for the functions $\tilde\Psi_{n,n\pm 1}(\vec
x+i\vec\delta).$ The zero value of $r_{n,n\pm 1}$, i.e. the zero value of the integral
$\int\biggl(\tilde\Psi^A_{n,n\pm 1}(\vec x+i\vec\delta)\biggr)^2d^2x$, definitely
signals {\bf incompleteness}
of the resolution of identity in terms of (nondegenerate) vectors $\tilde\Psi_{n,m},
\tilde\Psi^{\star}_{n,m}.$ In order to give a physical interpretation to the model, one
should complete the biorthogonal basis by suitable additional vectors.

Recently the problem of investigation of non-diagonalizable Hamiltonians in
one-dimensional Quantum Mechanics with non-Hermitian Hamiltonians was studied in papers
\cite{non} (see also the monographs \cite{naimark}). Up to our knowledge, not much
is known about two-dimensional non-diagonalizable Hamiltonians. One can conjecture that
the procedure to complete the basis is somehow similar to the one-dimensional case.
Then, in addition, one should consider (in the simplest case) the so called (first
order) associated functions $\tilde\Phi_{n-1,n}(\vec x+i\vec\delta),$ which solve the
inhomogeneous equation:
\be
(\tilde H - E_{n,n\pm 1})\tilde\Phi_{n,n\pm 1}=\tilde\Psi_{n,n\pm 1},
\label{inhomo}
\ee
where the function in r.h.s. is the normalizable eigenfunction of $\tilde H$ with
eigenvalue $E_{n,n\pm 1}.$

Then, due to the second Green's identity (Ostrogradsky-Gauss theorem),
the equalities ($\partial /\partial N$ - normal derivative)
\ba
&&0=\int\biggl(\tilde\Psi_{n,n\pm 1}(\vec x+i\vec\delta)\biggr)^2 d^2x=
\int\biggl((\tilde H-E_{n,n\pm 1})\tilde\Phi_{n,n\pm 1}\biggr)\tilde\Psi_{n,n\pm
1}d^2x=\nonumber\\
&&=\int\tilde\Phi_{n,n\pm 1}\biggl((\tilde H-E_{n,n\pm 1})\tilde\Psi_{n,n\pm
1}\biggr)d^2x-\oint_C(\tilde\Phi_{n,n\pm 1}\frac{\partial }{\partial N}\tilde\Psi_{n,n\pm
1})+\nonumber\\&&+\oint_C(\tilde\Psi_{n,n\pm 1}\frac{\partial}{\partial N}\tilde\Phi_{n,n\pm 1}),
\label{intt}
\ea
demonstrate that the integral over the large contour in the r.h.s. must be zero for
arbitrary solution $\tilde\Phi_{n,n\pm 1}$ of (\ref{inhomo}),
irrespectively of the fact that it is normalizable or not normalizable.

In one dimensional models with discrete spectrum (see [12]) for
the normalizable case one can complete the biorthogonal basis with
$\tilde\Phi_{n,n\pm 1}, \tilde\Phi^{\star}_{n,n\pm 1}$ with corresponding non-diagonal
terms in the resolution of identity. Then the Hamiltonian $\tilde H$ includes Jordan
blocks, and it is called non-diagonalizable.

In the two-dimensional case with discrete spectrum, the general discussion is rather
complicated. So, we restrict ourselves to the simplest case $n=0,m=1$ in order to
provide some analytical insight without ambition to propose general theorems.

In this particular case:
\be
(H-E_{0,1})\Phi^S_{0,1}=\Psi^S_{0,1};\quad E_{0,1}=-2\alpha^2s_0(s_0-1),
\label{in}
\ee
where the Hamiltonian with separation of variables is:
\ba
&&H=-\alpha^2\biggl(\xi_1^2\partial_1^2+\xi_2^2\partial_2^2+\xi_1\partial_1+\xi_2\partial_2-\frac{1}{4}(\xi_1^2+\xi_2^2)+(s_0+\frac{1}{2})(\xi_1+\xi_2)-1\biggr);\label{Hksi}\\
&&\xi_1=\frac{2\sqrt{A}}{\alpha}\exp{[-\alpha(x_1+i\delta)]};\quad
\xi_2=\frac{2\sqrt{A}}{\alpha}\exp{(-\alpha x_2)},
\nonumber
\ea
and the wave function reads:
\be
\Psi_{0,1}^S=\exp{[-\frac{\xi_1+\xi_2}{2}]}(\xi_1\xi_2)^{s_0}(\frac{1}{\xi_1}+\frac{1}{\xi_2}-\frac{2}{2s_0-1}).
\label{psiksi}
\ee
It is convenient to look for the particular solution $\Phi_{0,1}$ in the following form:
\be
\Phi^S_{0,1}=\exp{[-\frac{\xi_1+\xi_2}{2}]}(\xi_1\xi_2)^{s_0}
\bigl(\phi_1(\xi_1)+\phi_2(\xi_2)\bigr),
\label{phiksi}
\ee
where use has been made of separation of variables in Eq.(\ref{in}).
Correspondingly, one obtains that the function $\phi_1$ (and similarly for $\phi_2$)
fullfils an inhomogeneous ordinary differential equation:
\be
-\alpha^2\biggl(\xi_1^2\phi_1^{\prime\prime}-\xi_1^2\phi_1^{\prime}+(2s_0-1)\phi_1\biggr)=\frac{1}{\xi_1}-\frac{1}{2s_0-1}.
\label{Hksii}
\ee
The general solution of this equation can be expressed in terms of two linearly
independent solutions $y(\xi_1)$ and $z(\xi_1)$ with Wronskian $W$:
\be
\alpha^2\phi_1(\xi_1)=\mu y(\xi_1)+\nu
z(\xi_1)+z(\xi_1)\int^{\xi_1}_{0}d\tau\frac{y(\tau)(\frac{1}{\tau}
-\frac{1}{2s_0-1})}{\tau^2 W(\tau)}-
y(\xi_1)\int^{\xi_1}_{0}d\tau\frac{z(\tau)(\frac{1}{\tau}
-\frac{1}{2s_0-1})}{\tau^2 W(\tau)}.
\label{solution}
\ee
The analysis of asymptotic behaviour of $\phi_1$ and $\phi_2$ leads to the conclusion
that the function $\Phi^S_{0,1}$ is not normalizable and, in addition, the large
contour integral does not vanish. This is expected since the integral of $(\Psi_{0,1})^2$
is different from zero, the biorthogonal basis (\ref{ortho}) is complete, and the
Hamiltonian $H$ is diagonalizable.

Coming finally to the partner model with the Hamiltonian $\tilde H,$ we remind that
the integral of $(\tilde\Psi_{0,1})^2$ is zero. The partner (formal) associated function
$\tilde\Phi_{0,1}=Q^+\Phi_{0,1}$ turned out to be also nonnormalizable\footnote{It is necessary
to remind here again that $\tilde\Phi_{0,1}$ is only a particular solution of (\ref{inhomo}).}, however the
large contour integral vanishes
as required by (\ref{intt}). Therefore, the problem of completing of the resolution of
identity remains open.

The states $\tilde\Psi_0$ and $(\tilde\Psi_0)^{\star}$ must also be included in the
biorthogonal basis and the resolution of identity.
It is easy to show that the state $\tilde\Psi_0$ is orthogonal to $(\tilde\Psi_{n,m})^{\star}:$
\ba
&&\langle\biggl(\tilde\Psi_{n,m}(\vec x+i\vec\delta)\biggr)^{\star}\mid
\tilde{\Psi}_0(\vec x+i\vec\delta)\rangle
=\langle (Q^+)^{\star}\biggl(\Psi_{n,m}(\vec x+i\vec\delta)\biggr)^{\star}
\mid \tilde{\Psi}_0(\vec x+i\vec\delta)\rangle =\nonumber\\
&&=\langle\Psi^{\star}_{n,m}(\vec x+i\vec\delta)\mid Q^-
\tilde{\Psi}_0(\vec x+i\vec\delta)\rangle = 0.
\label{ooo}
\ea
It is difficult to find an analytic expression for the
pseudo-norm $\langle (\tilde{\Psi}_0)^{\star}\mid \tilde{\Psi}_0\rangle $ of the state (\ref{add})
but numerical evaluations
performed with positive values for the parameters
$s_0$ and $\delta$ varying in some limited range indicate that
the pseudo-norm does not vanish.

Summarizing, we have found that the biorthogonal
expansion related to Eq.(\ref{orthoo}) for $\tilde H$ is incomplete with appearance
of states $\tilde\Psi_{n,n\pm 1}$ of zero pseudo-norm. In one-dimensional
Quantum Mechanics this is associated to non-diagonalizability.
In our two-dimensional case we have not prooven  the existence of
associated
functions which are normalizable. Irrespectively of that we have
discovered an additional state for $\tilde H$ constructed from zero modes
of $Q^-$ which is pseudo-orthogonal to the other states. This vector definitely
should also enter the biorthogonal expansion for $\tilde H.$

Continuing the discussion of pseudo-Hermiticity, an imaginary coordinate shift
generates
this property for $\tilde H(\vec x + i\vec\delta)$, since the
following equation holds:
\begin{equation}\label{pseudo1}
  \tilde H(\vec x + i\vec\delta)=\tilde H^{\star}(\vec x - i\vec\delta)=
  \exp{(-2i\delta\partial )}\tilde H^{\dagger}(\vec x +
  i\vec\delta)\exp{(+2i\delta\partial )}.
\end{equation}
Comparing with (\ref{pseudo}), one can conclude that the explicit
form of operator $\eta$ in (\ref{pseudo}) can be written as: \be
\eta_{\delta} \equiv \exp{(+2i\vec\delta\vec\partial)}\equiv
O^{\dagger}O;\quad O\equiv \exp{(+i\vec\delta\vec\partial)}\equiv
O^{\dagger}. \label{pseudo2} \ee

In terms of $\eta$ (from now on the dependence on $\delta$ is not
written explicitly), the new (pseudo)inner product is defined
\cite{mostafazadeh} as:
\be
\langle\Omega (\vec x)|\Gamma (\vec x)\rangle_{\eta}
\equiv \langle\Omega (\vec x)|\eta \Gamma (\vec x)\rangle .
\label{product}
\ee
The precise form (\ref{pseudo2}) of $\eta$ gives for arbitrary $\Omega
(\vec x)$ and $\Gamma (\vec x)$: \be \langle\Omega (\vec x)|\Gamma (\vec x
)\rangle_{\eta} \equiv \langle\Omega (\vec x)|\eta \Gamma (\vec x)\rangle = \langle O\Omega (\vec
x)|O\Gamma (\vec x)\rangle = \int d^2x \biggl(\Omega (\vec
x+i\vec\delta)\biggr)^{\star} \Gamma(\vec x+i\vec\delta). \ee It is
clear now, why the pseudometric $\eta$ is positively definite: for $\Omega = \Gamma$
the $\eta -$norm is equal to
the integral of $|\Omega (\vec x+i\vec\delta)|^2$.

\section{\bf Conclusions.}
\vspace*{0.1cm}
\hspace*{3ex} Higher order (nonlinear) SUSY algebra has allowed us to
construct an {\bf exactly solvable} two-dimensional non-Hermitian
quantum model. We stress that this model {\it is not amenable} to
separation of variables, and it can be considered as a specific
$PT-$non-symmetric complexified version of generalized two-dimensional Morse model
with additional $\sinh^{-2}$ term. The spectrum of the model is real.
Here we focused attention on the
property of pseudo-Hermiticity of the model. To our knowledge this is
the first time that pseudo-Hermiticity is realized explicitly for a
nontrivial two-dimensional case. Following the general results, we
also studied the biorthogonal expansion and the metric operator
associated to pseudo-Hermiticity. In particular, it was shown that the Hamiltonian of
the model is not diagonalizable.

\section*{\bf Acknowledgments}
The work was partially supported by INFN, the University of
Bologna (M.V.I. and D.N.N.) and by the Russian grants
RFFI 06-01-00186-a, RNP 2.1.1.1112 (M.V.I.). M.V.I. is grateful to B.F.Samsonov and
A.V.Sokolov for useful clarifications of some statements about non-diagonalizable
Hamiltonians. 
\end{document}